\documentclass[aps,preprint,nofootinbib]{revtex4}
\usepackage{verbatim}   
\usepackage{bm}         
\usepackage{graphicx}   
\def\xp{{\bm x}^\perp}
\def\yp{{\bm y}^\perp}
\def\pp{{\bm p}_\perp}

\def\p{{\mathbf p}}
\def\k{{\mathbf k}}
\def\pip{{\bm \pi}_\perp}

\begin{document}
\vspace*{1cm}
\preprint {UW/PT 03-07}

\title{Construction of bosonic string theory
       on infinitely curved Anti-de~Sitter space}

\author {
  Adam Clark\footnotemark ,
  Andreas Karch\footnotemark ,
  Pavel Kovtun\footnotemark, and
  Daisuke Yamada\footnotemark}

\def\thefootnote{\fnsymbol{footnote}}
\footnotetext[1]{abc@u.washington.edu}
\footnotetext[2]{karch@phys.washington.edu}
\footnotetext[3]{pkovtun@u.washington.edu}
\footnotetext[4]{dyamada@u.washington.edu}
\def\thefootnote{\arabic{footnote}}
\affiliation
    {%
    Department of Physics,
    University of Washington,
    Seattle, Washington 98195-1560
    }%

\date{April 10, 2003\\[30pt]}

\begin{abstract}
Free scalar field theory in the sector with a large number of
particles can be interpreted as bosonic string theory on 
anti-de~Sitter space of vanishing radius. Different ways of writing the field
theory Hamiltonian translate to different ways of reparametrizing
the world-sheet $\sigma$ coordinate. Adding a mass term in the
field theory corresponds to cutting off the warped AdS direction,
with cut-off inversely proportional to the mass. The string theory
has neither tachyon, nor critical dimension.

\end{abstract}

\let\tpg=\thepage
\def\thepage{}          
\maketitle

\section{Introduction}
\let\thepage=\tpg       

The holographic AdS/CFT duality conjecture states the equivalence
between string/M theories on various background spaces, and gauge
theories, formulated on the boundary of
those spaces%
~\cite{AdSCFT Review}. In the context of the duality, a
correspondence exists between important dimensionless parameters,
which control the dynamics. A string theory has two such
parameters. One is the ratio of string tension, $T=1/(2\pi\alpha')$,
to the characteristic curvature of the background space. Another
is the string coupling constant, which sets the probability of string
splitting and joining. On the field theory side, these parameters
correspond to the 't~Hooft coupling and to the number of colors $N$
in gauge theory. Thus, the $1/N$ expansion in gauge theory corresponds
to the genus expansion of string world-sheets, while finite 't~Hooft
coupling effects correspond to stringy corrections to the
effective low-energy string theory. In particular, field theories
at infinite 't~Hooft coupling are dual to theories of classical
supergravity.

Following the duality recipe, field theories at small coupling
must correspond to string theories formulated on backgrounds
whose characteristic radius of curvature is much smaller than
the string length. Since geometry is supposed to be dynamical in
string theory, it is not clear what the duality means in this
weak-coupling regime. The problem is, of course, that string theory
lacks a
(non-perturbative) definition.
Turning things around, one can choose to use the duality to
\emph{define} string theory in the limit of small tension. It is
an important unresolved problem how to construct a general
definition.

In this note we consider bosonic string theory on the background
of Anti-de Sitter space (rather, its Poincare patch), whose radius
of curvature $R$ is exactly zero. In this limit, the classical string
Hamiltonian simplifies considerably, and it is not difficult
to quantize the theory in the light-cone gauge.%
\footnote{
  Such simplification occurs
  only because the limit $R^2/\alpha' \to 0$
  (which gets rid of certain \emph{derivative} terms)
  is taken in the classical string Hamiltonian.
  In general, we do not expect that the same limit, when taken in
  the quantum Hamiltonian (whatever it is) will produce
  the same result.
  In other words, there is no good reason to believe 
  that the zero-radius solution
  is the correct starting point to define string theory
  perturbatively in $R^2/\alpha'$.
  For attempts to construct a systematic
  large-curvature expansion in string theory, see
  \cite{large curvature}.
  }
What we find is a very natural correspondence between string
theory in this singular limit and a free matrix-valued scalar
field theory in the sector of single-trace states with a large
number of particles. The correspondence
was first noted in \cite{Andreas}, but the interpretation there
was not clear. One can make several observations about the
correspondence:
\begin{itemize}
\item Both Hamiltonians and states are easily mapped between
string and field theory.

\item Part of the world-sheet reparametrization invariance of string theory
corresponds to the freedom to define different continuum limits in
the field theory.

\item The field theory Hamiltonian provides a discretization of
the string theory Hamiltonian. For states with a large number of
particles, the jagged string world-sheet is a good approximation.

\item By checking the closure of the Lorentz algebra in string
theory, one finds that there is no critical dimension. This is in
natural correspondence with the fact that free field theories
exist in any number of spacetime dimensions.

\item The spectrum of the string Hamiltonian does not contain a tachyon,
naturally reflecting the stability of the field theory.

\item Adding mass $m$ to the scalar particles corresponds to
imposing a hard cutoff equal to $2\pi/m$ 
in the warped direction of the Poincare
patch. This is very similar to the standard (super)gravity
construction of confining gauge theories (in the large 't~Hooft
coupling limit of the AdS/CFT correspondence). Mode number in
the bulk maps to particle number on the boundary.

\item The large $N$ expansion in the field theory (which in the absence of
interactions becomes just the combinatorics of organizing Wick
contractions) is mapped to the ``genus expansion'' in string theory.

\end{itemize}

The very idea of explicitly building string theories from large-$N$
field theories, possibly with some sort
of discretization, is rather old 
\cite{t Hooft, Giles Thorn, Klebanov Susskind, Dalley Klebanov}.
Recently, significant progress has been made
in constructing string world-sheets
from large-$N$ planar diagrams
\cite{Bardakci Thorn}. In the approach we take in this note,
we are not trying to construct a
formulation which specifically requires smooth world-sheet
embeddings. The string theory we are discussing has essentially
zero tension, and therefore neighboring bits on the string have
no correlation in space-time. Rather, our goal is 
to make
a connection between free scalar field theory, and the naive
small-radius limit of string theory on Anti-de Sitter space. 
The paper is organized as follows. 
In the next section we review 
quantization of a free scalar 
field on the light-front, and 
show how, for single-trace states,
a picture of a discretized string 
emerges. 
We also discuss different continuum
limits for the free theory Hamiltonian
in the sector with a large number
of particles. 
In Section~\ref{sec:string} we show
that the same Hamiltonian arises as
the light-cone Hamiltonian of bosonic
string theory on zero radius AdS space
with a hard cut-off on the warped dimension.
The number of particles is interpreted
as the discrete mode number of the 
warped dimension. 
In Section~\ref{sec:critical dimension}
we check that the string theory on
zero radius AdS space passes the most
crucial test a consistent string theory
in light cone gauge has to pass:
the Lorentz anomaly vanishes, independent
of the number of dimensions. 
In Section~\ref{sec:discussion}
we conclude and discuss
directions for future work.

\section{Free scalar field theory on the light-cone}

We start with the following action for a free scalar
field theory in flat Minkowski space:
\begin{equation}
   S = \int d^{\,d} x \left(
       -\frac 12 {\rm tr}(\partial \Phi)^2
       -\frac 12 m^2 \, {\rm tr} \Phi^2 \right) \ .
\end{equation}
Here $\Phi\equiv\Phi_{ij}(x^0,x^1,\dots,x^{d-1})$
is an $N \times N$ hermitian matrix field,
and we use the mostly plus metric.
The action has a $U(N)$ symmetry,
which transforms $\Phi\to U \Phi U^\dagger$.
The theory can be quantized on the light-front,
as an alternative to equal-time canonical
quantization, see e.g. \cite{Chang Root Yan, LC review}.
To do so, one introduces light-cone coordinates
$x^{\pm} = (x^1 \pm x^0)/\sqrt{2}$,
and commutation relations for fields
$\Phi(x^+,x^-,x^2,\dots,x^{d-1}) \equiv
 \Phi(x^+,x^-,\xp)$
are imposed at equal $x^+$,
rather than at equal $x^0$:
\begin{equation}
    \left[ \partial_- \Phi_{ij}(x^+, x^-, \xp),
    \Phi_{kl}(x^+, y^{\,-},\yp) \right] =
    - \, \delta_{ik}  \delta_{jl} \; \frac i2 \,
    \delta(x^- - y^{\,-}) \, \delta(\xp - \yp)
\end{equation}
where $\partial_- \equiv \partial/\partial x^-$.
Mode expansions for ``Schr\"odinger picture'' operators
now become
\begin{equation}
    \Phi_{ij}(x^-,\xp) =
    \int_0^\infty \frac{dp_-}{2\pi 2p_-}
    \int \frac{d^{\,d-2}p_\perp}{(2\pi)^{d-2}}  \left[
    a_{ij}(p_-,\pp) e^{ ip_- x^- + i\pp\cdot\xp} +
    a^\dagger_{ij}(p_-,\pp) e^{ -ip_- x^- - i\pp\cdot\xp}
    \right]
\end{equation}
and the light-cone Hamiltonian $H_{LC}$
generates translations in $x^+$.
The commutation relations for the creation/annihilation
operators are
\begin{eqnarray}
    &&[a_{ij}(\p), a^\dagger_{kl}(\k)] =
      \delta_{ik} \delta_{jl} 2 p_- (2\pi)^{d-1}
      \delta(\p-\k) \\
    &&[a_{ij}(\p), a_{kl}(\k)] =
      [a^\dagger_{ij}(\p), a^\dagger_{kl}(\k)] = 0 \ .
\end{eqnarray}
The vacuum state $|0\rangle$ is annihilated
by all $a_{ij}(\p)$, and a general
$M$-particle state is
\begin{equation}
    a^\dagger_{i_1 j_1}(\p_1) \,
    a^\dagger_{i_2 j_2}(\p_2) \dots
    a^\dagger_{i_M j_M}(\p_M) \,|0\rangle \ ,
\end{equation}
where we use $\p$ as short for $(p_-,\pp)$.
The single-particle dispersion relation takes the form
\begin{equation}
    {\cal E}_{LC}(\p) = - p_+
    = \frac{\pp^2 + m^2}{2 p_-}
\end{equation}
and thus positivity of energy requires $p_->0$.
For states with a given number $M$ of particles,
the light-cone Hamiltonian takes the usual
second-quantized form:
\begin{equation}
\label{eq:M particle Hamiltonian 1}
    H_{LC} = \sum_{\p} N_\p \;
             {\cal E}_{LC}(\p)
           = \sum_{\p} N_\p \;
             \frac{\pp^2 + m^2}{2 p_-} \ ,
\end{equation}
where $N_\p$ is the number operator,
and $\sum_{\p} N_\p = M$.

\subsection{Continuum limit with constant density of particles}

Instead of summing over momenta,
the same Hamiltonian can be
rewritten as a sum over particles.
Knowing the number of particles $N_\p$
with given momentum, one can assign
momentum $\p(n)$ to every integer $n$
between $1$ and $M$.
Such an assignment is, of course,
not unique, because the labeling
of particles is arbitrary.
Then the Hamiltonian
(\ref{eq:M particle Hamiltonian 1})
becomes
\begin{equation}
\label{eq:M particle Hamiltonian 2}
     H_{LC} = \sum_{n=1}^M
     \frac{\pp^2 (n) + m^2}{2 p_- (n)} \ ,
\end{equation}
where $p_- (n)$, $\pp (n)$ are
momenta of the $n$-th particle.
A natural assignment of $\p(n)$
can be made, when one considers
single-trace $M$-particle states:%
\footnote{
     Restricting to the subsector of 
     single-trace states is consistent
     at large $N$, when the overlap of 
     single-trace and multi-trace states
     is suppressed by powers of $1/N$. 
     As usual, single-trace states will be
     interpreted as single-string states, 
     multi-traces as multi-string states,
     and at large $N$ splitting and joining
     string interactions do not happen.}
\begin{eqnarray}
   && |\Psi^M \rangle \sim {\rm tr} \left[
      a^\dagger(\p_1)
      a^\dagger(\p_2) \dots
      a^\dagger(\p_M) \right] \,|0\rangle \\
   && \sum_{n=1}^M p_{n-} = P_- \ , \ \ \ \
      \sum_{n=1}^M {\bm p}_{n\perp} = {\bm P}_\perp
\end{eqnarray}
For such states, there is a natural
way to label the particles, up to
cyclic permutations.
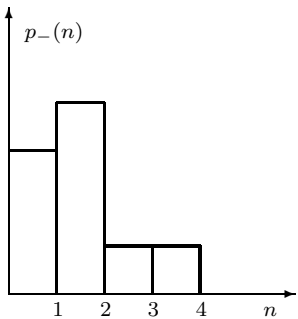
\begin{figure}
\setlength{\unitlength}{3947sp}%

\begin{picture}(2154,2154)(859,-1603)
\thinlines
{\put(901,-1561){\vector( 0, 1){1800}}
}%
\thicklines
{\put(901,-661){\line( 1, 0){300}}
\put(1201,-661){\line( 0, 1){300}}
\put(1201,-361){\line( 1, 0){300}}
\put(1501,-361){\line( 0,-1){900}}
\put(1501,-1261){\line( 1, 0){300}}
\put(1801,-1261){\line( 1, 0){300}}
\put(2101,-1261){\line( 0,-1){300}}
}%
\thinlines
{\put(901,-1561){\vector( 1, 0){1800}}}
\thicklines
{\put(1201,-661){\line( 0,-1){900}}}
{\put(1501,-1261){\line( 0,-1){300}}}
{\put(1801,-1261){\line( 0,-1){300}}}
%
{\put(1175,-1700){${\scriptstyle 1}$}}
{\put(1475,-1700){${\scriptstyle 2}$}}
{\put(1775,-1700){${\scriptstyle 3}$}}
{\put(2075,-1700){${\scriptstyle 4}$}}
{\put(2500,-1700){${\scriptstyle n}$}}
{\put(1000,50){${\scriptstyle p_-(n)}$}}
\end{picture}

\caption{
         $M=4$ particles with total momentum
         $P_- = \sum_{n=1}^M p_-(n)$}
\label{fig:constant particle density}
\end{figure}
The states carry definite momenta,
and can be labeled as
\begin{equation}
     |(p_-,\pp)_{n_1}, (p_-,\pp)_{n_2}, \dots
     (p_-,\pp)_{n_M} \rangle \ .
\end{equation}
In other words, one can think of
these states as strings of particles,
created by the string of operators
inside the trace.
One way to visualize these states is shown
in Fig.~\ref{fig:constant particle density}.
When the number of particles $M$ is
large, the summation over particle
number in the Hamiltonian
($\ref{eq:M particle Hamiltonian 2}$)
can be represented as an integral over
positions of the particles:
\begin{equation}
\label{eq:M particle Hamiltonian 3}
    H_{LC}= \sum_{n=1}^M
    \frac{\pp^2 (n) + m^2}{2 \, p_- (n)}
    \to
    \frac 1a \int_0^l dx\;
    \frac{\pp^2 (x) + m^2}{2 p_- (x)}
\end{equation}
where $a \equiv l/M$ is a constant
``lattice spacing'', and $1/a$ is
particle density on the string.
The cyclic symmetry of the trace implies
that $p_-(x)$ and $\pp(x)$ are periodic
functions of $x$ on the interval $(0,l)$.
In other words, the string of particles
is closed.%
\footnote{
    Open strings of particles naturally appear
    when the theory contains fields, which
    transform as vectors under the $U(N)$
    symmetry. If $b^\dagger_i$ are creation
    operators for such fields, then open string
    states have the form
    $|\Psi_M\rangle \sim
    b^\dagger_{j_1}(\p_1) a^\dagger_{j_1 j_2}(\p_2)
    a^\dagger_{j_2 j_3}(\p_3) \dots
    a^\dagger_{j_{M-2}j_{M-1}}(\p_{M-1})
    b^\dagger_{j_{M-1}}(\p_M)
    |0\rangle$.
    }
The functions $p_-(x)$ and $\pp(x)$
are integrable, but not smooth,
because particle momenta in
$|\Psi^M\rangle$ are arbitrary.
One can rewrite the Hamiltonian
(\ref{eq:M particle Hamiltonian 3})
in terms of momentum densities
$\pi_-(x) \equiv p_- (x)/a$,
$\pip(x) \equiv \pp(x)/a$,
$\pi_Y \equiv m/a$
on the string:
\begin{equation}
\label{eq:FT continuum Hamiltonian 1}
    H_{LC} = \int_0^l dx\;
    \frac{\pip^2 (x) + \pi_Y^2}{2 \, \pi_- (x)}
\end{equation}
Momentum densities
are functions of $x$, but particle
density on the string is
$x$-independent.

\subsection{Continuum limit with constant momentum density}

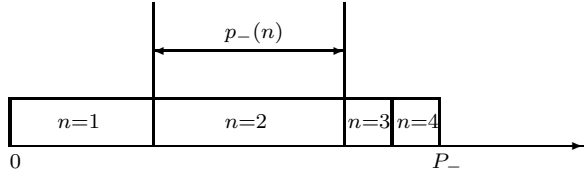
\begin{figure}

\begin{center}
\setlength{\unitlength}{3947sp}%
\begin{picture}(3634,954)(879,-703)

\thicklines
{\put(901,-661){\line( 0, 1){300}}
\put(901,-361){\line( 1, 0){900}}
\put(1801,-361){\line( 0,-1){300}}
\put(1801,-661){\line( 0, 1){300}}
\put(1801,-361){\line( 0,-1){300}}}
{\put(1801,-361){\line( 1, 0){1200}}
\put(3001,-361){\line( 0,-1){300}}}
{\put(3001,-361){\line( 1, 0){300}}
\put(3301,-361){\line( 0,-1){300}}}
{\put(3301,-361){\line( 1, 0){300}}
\put(3601,-361){\line( 0,-1){300}}}
\thinlines
{\put(1801,239){\line( 0,-1){600}}}
{\put(3001,239){\line( 0,-1){300}}
\put(3001,-61){\line( 0,-1){300}}}
{\put(1801,-61){\vector(-1, 0){  0}}
\put(1801,-61){\vector( 1, 0){1200}}}
{\put(901,-661){\vector( 1, 0){3600}}}
{\put(2250,20) {${\scriptstyle p_- (n)}$}}
{\put(901,-800) {${\scriptstyle 0}$}}
{\put(3550,-800) {${\scriptstyle P_-}$}}
{\put(1200,-550) {${\scriptstyle n=1}$}}
{\put(2250,-550) {${\scriptstyle n=2}$}}
{\put(3025,-550) {${\scriptstyle n=3}$}}
{\put(3330,-550) {${\scriptstyle n=4}$}}

\end{picture}
\end{center}
\caption{
         The same $M=4$ particles with total
         momentum $P_- = \sum_{n=1}^M p_-(n)$}
\label{fig:constant momentum density}
\end{figure}

One can look at the same multiparticle
states from another point of view.
With the natural ordering in the
single-trace states, particles can be
arranged on the momentum axis.
A simple example is shown in
Fig.~\ref{fig:constant momentum density}.
For a given total momentum $P_-$ of the state,
when the number of particles $M$ becomes
large, the summation over particle
number in the Hamiltonian
($\ref{eq:M particle Hamiltonian 2}$)
can be represented as an integral over
momenta of the particles:
\begin{equation}
\label{eq:M particle Hamiltonian 4}
    H_{LC}= \sum_{n=1}^M
    \frac{\pp^2 (n) + m^2}{2 \, p_- (n)}
    \to
    \int_0^l \frac{dx}{\tilde a(x)}
    \frac{\pp^2 (x) + m^2}{2 p_-(x)} =
    \int_0^l dx\; \frac{1}{\tilde a(x)^2}
    \frac{\pp^2 (x) + m^2}{2 k_-} \ .
\end{equation}
Here we chose the position-dependent
``lattice spacing'' to be
$\tilde a = p_-(n)/k_- \to p_-(x)/k_-$,
where $k_- = P_-/l$ is a constant,%
\footnote{
     It is convenient to choose the arbitrary
     length $l$ to be the same in
     (\ref{eq:M particle Hamiltonian 4}) and
     (\ref{eq:M particle Hamiltonian 3}).
     }
whose physical meaning is $p_-$-momentum
density on the string.
One can rewrite the Hamiltonian
(\ref{eq:M particle Hamiltonian 4})
in terms of momentum densities
$\tilde\pi_Y(x) \equiv m/\tilde a(x)$,
$\tilde\pip(x) \equiv \pp(x)/\tilde a(x)$
on the string:
\begin{equation}
\label{eq:FT continuum Hamiltonian 2}
    H_{LC} = \frac{1}{2k_-} \int_0^l dx\;
    \left( \tilde\pip^2 (x) + \tilde\pi_Y^2(x) \right)
\end{equation}
In this approach, $p_-$-momentum density
is constant, but particle density on the
string, $1/\tilde a$, is a function of $x$.

\section{String theory on Anti-de Sitter space of vanishing radius}
\label{sec:string}

\subsection{Classical string theory on a given background}

To maintain self-consistency
of our presentation, we derive
in this subsection the classical
light-cone Hamiltonian for bosonic
string theory on AdS space
\cite{Polchinski Susskind, Metsaev Thorn Tseytlin}.
One starts with the Polyakov action
for the bosonic string on a general $(d+1)$
dimensional background space:
\begin{equation}
\label{eq:string action}
  S=-\frac{1}{4\pi\alpha'}
       \int d\tau \int_0^l d\sigma
       \sqrt{-\gamma}\gamma^{ab} \,
       \partial_a Z^\mu \partial_b Z^\nu \, G_{\mu\nu}(Z) \ .
\end{equation}
Indices $a,b$ label world-sheet
coordinates $\tau,\sigma$, and
indices $\mu,\nu$ run from 0 to $d$.
$G_{\mu\nu}(Z)$ is the background
metric with signature $-+...+$, and
$\gamma_{ab}$ is the world-sheet metric.
It will be convenient to make use of
$h^{ab} = \sqrt{-\gamma} \gamma^{ab}$,
which satisfies the identity
$(h^{\tau\sigma})^2 - h^{\tau\tau} h^{\sigma\sigma}=1$.
%
%
Canonical momentum densities for the coordinates $Z^\mu$,
\begin{equation} \label{eq:canon-momenta}
\pi_\mu = \frac{\delta\mathcal{L}}{\delta \dot Z^\mu} =
          -\frac{1}{2\pi\alpha'}G_{\mu\nu} \left(
          h^{\tau\tau} \dot Z^\nu + h^{\tau\sigma} Z^{\nu\prime}
          \right) \ ,
\end{equation}
give rise to the following Hamiltonian density:
\begin{equation}
\mathcal{H} = -\frac{\pi\alpha'}{h^{\tau\tau}} \left[
              G^{\mu\nu}\pi_\mu \pi_\nu +
              \frac{1}{(2\pi\alpha')^2}
              G_{\mu\nu}Z^{\mu\prime} Z^{\nu\prime}
              \right] -
              \frac{h^{\tau\sigma}}{h^{\tau\tau}}
              \pi_\mu Z^{\mu\prime} \ ,
\end{equation}
where dot and prime denote derivatives with respect
to $\tau$ and $\sigma$, respectively.
Components of the world-sheet metric are non-dynamical fields,
and therefore the equations of motion for
$h^{\tau\tau}$ and $h^{\tau\sigma}$
become constraint equations for the coordinates $Z^{\mu}$:
\begin{eqnarray}
    &&G^{\mu\nu}\pi_\mu \pi_\nu +
    \frac{1}{(2\pi\alpha')^2}
    G_{\mu\nu}Z^{\mu\prime} Z^{\nu\prime} = 0,
    \label{eq:constraint1}\\
    &&\pi_\mu Z^{\mu\prime} = 0.
    \label{eq:constraint2}
\end{eqnarray}

It is convenient to introduce light-cone coordinates
\begin{equation}
Z^\pm = \frac{1}{\sqrt{2}}\left(Z^1\pm Z^0\right) \ .
\end{equation}
We will label the remaining $Z^\mu$ coordinates by small
Latin indices $i,j,...$ which run from 2 to $d$.
Quantities with lower $\pm$ indices are defined similarly.%
\footnote{
     For the metric, which lowers $\pm$ indices, one finds
     $G_{++}=\frac{1}{2}\left(G_{00}+2G_{01}+G_{11}\right)$,
     $G_{+-}=G_{-+}=\frac{1}{2}\left(G_{11}-G_{00}\right)$,
     $G_{--}=\frac{1}{2}\left(G_{00}-2G_{01}+G_{11}\right)$,
     $G_{\pm i}=G_{i\pm}=\frac{1}{\sqrt{2}}\left(G_{1i} \pm G_{0i}\right)$.
     Similar relations hold for $G^{++}$, $G^{--}$, and $G^{+-}$.}
We will take the background space to be such that
$G_{\pm\,i}=G_{++}=G_{--}=0$,
which implies $G^{+-}=(G_{+-})^{-1}$.

To proceed with the dynamics of the fields $Z^\mu$,
we go to the light-cone gauge:
one can use $\tau,\sigma$ reparametrization invariance
of $S_P$ to fix
\begin{eqnarray}
   Z^+=\tau  \ ,        \label{eq:gaugefix1}\\
   h^{\tau\sigma}=0 \ . \label{eq:gaugefix2}
\end{eqnarray}
Using the gauge-fixing conditions
(\ref{eq:gaugefix1}), (\ref{eq:gaugefix2}),
the Lagrangian density becomes
\begin{equation} \label{eq:LC Lagrangian}
    \mathcal{L}_{LC}=-\frac{h^{\tau\tau}}{4\pi\alpha'} \left(
    2 G_{+-} \dot Z^- +  G_{ij} \dot Z^i \dot Z^j -
    \frac{1}{(h^{\tau\tau})^2} G_{ij} Z^{i\,\prime} Z^{j\,\prime}
    \right)
\end{equation}
The equations of motion which follow from this
Lagrangian are to be supplemented by the constraint
equations (\ref{eq:constraint1}), (\ref{eq:constraint2}),
which in the light-cone gauge read
\begin{eqnarray}
    &&2\,G^{+-}\pi_+ \pi_- + G^{ij}\pi_i \pi_j +
    \frac{1}{(2\pi\alpha')^2} G_{ij} Z^{i\,\prime} Z^{j\,\prime} =0
    \label{eq:LC-constraint1} \\
    &&\pi_- Z^{-\prime} + \pi_i Z^{i\,\prime} = 0 \ ,
    \label{eq:LC-constraint2}
\end{eqnarray}
where $\pi_+$, $\pi_-$ are conjugate momenta for
the fields $Z^+$, $Z^-$.
The equation of motion for $Z^-$ implies that
its canonical momentum
is a function of $\sigma$ only: $\pi_-=\pi_-(\sigma)$.
The single-string light-cone Hamiltonian,
which one obtains from the Lagrangian
(\ref{eq:LC Lagrangian}), is
\begin{equation}\label{eq:LC Hamiltonian}
    H_{LC} = \int_0^l \mathcal{H}_{LC} =
    \int_0^l d\sigma \, \frac{G_{+-}}{2 \,\pi_-}
    \left( G^{ij} \pi_i \pi_j +
    \frac{1}{(2\pi\alpha')^2} G_{ij} Z^{i\,\prime} Z^{j\,\prime}
    \right)
\end{equation}
and using the constraint (\ref{eq:LC-constraint1}),
it is easy to see that $\mathcal{H}_{LC}$ is just the
negative of $\pi_+$:
\begin{equation}
  \mathcal{H}_{LC}=-G_{+-}G^{-+}\pi_+ = -\pi_+ \ .
\end{equation}

This form of the Hamiltonian is very convenient,
when the background space is $(d+1)$-dimensional
Anti-de Sitter space in Poincare coordinates:
\begin{equation}
\label{eq:AdS metric}
    ds^2_{AdS}=\frac{R^2}{Y^2} \left(
    2 dZ^+ dZ^- + dZ^i dZ^j \delta_{ij}
    \right).
\end{equation}
Here $R$ is ($Z^\mu$-independent) curvature radius
of the space, and $Y \equiv Z^{\mu=d}$.
The ``warped coordinate'' $Y$ must take non-negative values,
while the rest of the $Z^\mu$ can be of any sign.
The boundary of the space is at $Y=0$.
With the AdS metric (\ref{eq:AdS metric}),
the light-cone Hamiltonian (\ref{eq:LC Hamiltonian})
takes the simple form
\begin{equation}\label{eq:LC AdS Hamiltonian}
   H_{LC}^{(AdS)}= \int_0^l \frac{d\sigma}{2\, \pi_-} \, \left(
                     \pi_i \pi_i +
                     \frac{1}{(2\pi\alpha')^2}
                     \frac{R^4}{Y^4} Z^{i\,\prime} Z^{i\,\prime}
                     \right)\ .
\end{equation}

\subsection{The choice of $\sigma$ coordinate}
\label{sec:choice of sigma}

The Hamiltonian (\ref{eq:LC AdS Hamiltonian})
can be further simplified
by using some leftover reparametrization invariance --
the reparametrizations of $\sigma$ only.
One convenient way to fix the $\sigma$ coordinate
is to measure it in units of $\pi_-(\sigma)$,
while keeping the total string length $l$ fixed.
This is a natural choice, because
$\pi_-(\sigma) \sim -h^{\tau\tau} =
\gamma_{\sigma\sigma}/\sqrt{-\gamma}$,
and therefore it transforms under reparametrizations
$\sigma\to\tilde\sigma(\sigma)$ as
\begin{equation} \label{eq:pi- transformation}
    \tilde\pi_-(\tilde\sigma)d\tilde\sigma = \pi_-(\sigma)d\sigma \ .
\end{equation}
Let us choose the new $\sigma$ coordinate as%
\footnote{
    $\pi_- = -h^{\tau\tau}G_{+-}/(2\pi\alpha')$ is positive
    because both $\gamma_{\sigma\sigma}$ and $G_{+-}$ are
    positive. Therefore, changing $\sigma$ to $\tilde\sigma$
    involves no back-tracking.}
\begin{equation}
   \tilde\sigma = \frac{1}{p_-}
   \int_0^{\sigma} \pi_-(\sigma') d\sigma' \ ,
\end{equation}
where
$p_- \equiv \frac{1}{l} \int_0^l \pi_-(\sigma') d\sigma'$.
We will refer to this choice of $\sigma$
as the $\pi_- = const$ gauge.
Then the light-cone Hamiltonian
(\ref{eq:LC AdS Hamiltonian}) becomes
\begin{equation}\label{eq:LC AdS Hamiltonian 1}
   H_{LC}^{(AdS)}= \frac{1}{2 p_-}\int_0^l d\sigma \, \left(
                     \pi_i \pi_i +
                     \frac{1}{(2\pi\alpha')^2}
                     \frac{R^4}{Y^4} Z^{i\,\prime} Z^{i\,\prime}
                     \right)\ .
\end{equation}
This Hamiltonian can be compared
with the Hamiltonian of string
theory on a flat $d+1$ dimensional
background:
\begin{equation}
\label{eq:flat Hamiltonian}
    H_{LC}^{(flat)} =
    \frac{1}{2 p_-}\int_0^l d\sigma \, \left(
    \pi_i \pi_i +
    \frac{1}{(2\pi\alpha')^2}
    Z^{i\,\prime} Z^{i\,\prime}
                     \right)\ .
\end{equation}
Thus, strings in Anti-de Sitter space
can be thought of as having a variable tension,
$\frac{1}{2\pi\alpha'}\frac{R^2}{Y^2}$,
which increases as one approaches
the boundary of the space.

When the radius of curvature
is exactly zero, 
the AdS Hamiltonian takes the simple form%
\footnote{
    When the radius of curvature is non-zero, the second
    term in the Hamiltonian (\ref{eq:LC AdS Hamiltonian 1})
    becomes important as one approaches the boundary at $Y=0$.
    The second term also contains derivatives of the fields
    $Z^i$, and therefore can not in general be treated
    as a small perturbation, even when $R^2/\alpha'$ is small.}
\begin{equation}
\label{eq:LC zeroAdS Hamiltonian}
    H_{LC} = \frac{1}{2 p_-}\int_0^l d\sigma \;
    \pi_i \pi_i   \equiv
    \frac{1}{2 p_-}\int_0^l d\sigma \;
    \left( \bm{\pi}_{\perp}(\sigma)^2 +
    \pi_Y(\sigma)^2\right) \ ,
\end{equation}
where $\pi_Y\equiv\pi_{i=d}$,
and momentum densities are written as
$\pi_i=(\bm{\pi}_\perp, \pi_Y)$.
This limit corresponds to vanishing tension,
and therefore describes the situation when
each piece of the string moves independently.
The Hamiltonian (\ref{eq:LC zeroAdS Hamiltonian})
is precisely the Hamiltonian
(\ref{eq:FT continuum Hamiltonian 2})
of the free field theory, which describes
a collection of free particles.
A particle in field theory corresponds to a
freely moving bit of a tensionless string.

When $R^2/\alpha'=0$, other choices of
parametrizing $\sigma$ are also useful.
Namely, when the radius is zero, equations of
motion require all transverse
momenta $\pi_i$ to be $\tau$-independent.
All of them are proportional to $h^{\tau\tau}$,
and therefore change under reparametrizations
just like $\pi_-$ in  (\ref{eq:pi- transformation}).
Measuring the $\sigma$ coordinate
in units of $|\pi_Y(\sigma)|$ will be
particularly convenient;
in this case the Hamiltonian becomes
\begin{equation}
\label{eq:py string Hamiltonian}
   H_{LC} = \int_0^l d\sigma \, \frac{
                     \sum_{i=2}^{d-1} \pi_i \pi_i + p_Y^2}
                     {2 \, \pi_-(\sigma)} \equiv
           \int_0^l d\sigma \, \frac{
                     \bm{\pi}_{\perp}(\sigma)^2 + p_Y^2}
                     {2 \, \pi_-(\sigma)}\ ,
\end{equation}
where $p_Y=\frac{1}{l}\int_0^l |\pi_Y (\sigma)| d\sigma$
is $\sigma$-independent.
Again, this is precisely the
field theory Hamiltonian
(\ref{eq:FT continuum Hamiltonian 1}),
when the continuum limit is taken
with constant density of particles.
This choice of parametrization will be
referred to as $\pi_Y = const$ gauge.
For the rest of the paper, we restrict ourselves
to the case $R^2/\alpha'=0$.

\subsection{Mode expansions}

The description of the dynamics becomes particularly
simple when distance along the string is
measured in units of $\pi_-(\sigma)$.
Canonical momenta become
$\pi_+ = p_- \dot Z^-$, $\pi_i = p_- \dot Z^{\,i}$,
and canonical equations of motion reduce to
$\ddot Z^{\,i} (\sigma,\tau) = 0$.
The independent degrees of freedom are
$Z^i$ and $\pi_i\ $; once $Z^{\,i}$
are known, one can determine
$Z^{-}(\sigma,\tau)$ (up to a constant)
from the constraints
(\ref{eq:LC-constraint1}), (\ref{eq:LC-constraint2}):
\begin{eqnarray}
    &&\dot Z^- = -\frac{1}{2p_-^2} \, \pi_i \pi_i  \ ,
    \label{eq:Zminusdot}\\
    &&Z^{-\prime} = -\frac{1}{p_-}\, \pi_i \, Z^{i\,\prime} \ .
    \label{eq:Zminusprime}
\end{eqnarray}
Since the transverse momenta are conserved,
$\dot \pi_i = 0$, the first of the constraints
implies that $Z^-$ satisfies the same equation as
the transverse coordinates: $\ddot Z^-(\sigma,\tau) = 0$.

A general solution to the equations of motion,
which satisfies Neumann boundary conditions
$Z^{\,i\,\prime}(\sigma=0,\tau) = 0 = Z^{\,i\,\prime}(\sigma=l,\tau)$
(open string with free ends) can be written as
\begin{eqnarray}
\label{eq:AdS-Open-Mode}
   &&Z^i (\sigma,\tau) = \sum_{n=-\infty}^{\infty} \left(
                        z^i_n + \frac{p^i_n}{l} \tau \right)
                        \cos\left(\frac{\pi n \sigma}{l}\right) \\
   &&\pi_i = p_- \dot Z^i \; = \; p_- \!\sum_{n=-\infty}^{\infty}
                        \frac{p^i_n}{l}
                        \cos\left(\frac{\pi n \sigma}{l}\right)
\end{eqnarray}
where $z^i_n = z^i_{-n}$, $p^i_n = p^i_{-n}$.
From (\ref{eq:Zminusprime}) it follows
that, if $Z^i$ satisfies Neumann boundary conditions,
then so does $Z^-$. This tells us that $Z^-$ has a mode expansion,
which is identical in form to the expansion of the transverse coordinates:
\begin{equation}
    Z^- (\sigma,\tau) = \sum_{n=-\infty}^{\infty} \left(
                        z_n^- + \frac{p_n^-}{l} \tau \right)
                        \cos\left(\frac{\pi n \sigma}{l}\right)
\end{equation}
Comparing this expansion with (\ref{eq:Zminusdot}),
(\ref{eq:Zminusprime}) determines $z_n^-$ and $p_n^-$ to be%
\footnote{The identity
          $\sum_{n=-\infty}^{\infty} \left( \frac{k}{2}-n \right)
          p^i_{k-n} p^i_n = 0$
          is of help.}
\begin{eqnarray}
    &&z_n^- = -\frac{1}{l} \sum_{m=-\infty}^{\infty}
             z^i_{n-m} \, p^i_m \left(1-\frac{m}{n}\right)
             \label{eq:small-z-minus} \\
    &&p_n^- = -\frac{1}{2l} \sum_{m=-\infty}^{\infty}
             p^i_{n-m} \, p^i_m
             \label{eq:small-p-minus}
\end{eqnarray}
The coefficient $z_0^-$ is left undetermined, which corresponds
to the fact that the constraints (\ref{eq:Zminusdot}),
(\ref{eq:Zminusprime}) determine $Z^{-}(\sigma,\tau)$
only up to an overall additive constant.
The rest of the mode expansion coefficients satisfy
 $z_{-n}^- = z_{n}^-$, $p_{-n}^- = p_{n}^-$.
The light-cone Hamiltonian (\ref{eq:LC zeroAdS Hamiltonian})
can be expressed in terms of the mode expansion coefficients:
\begin{equation}
\label{eq:open string H mode expansion}
     H_{LC} = \frac{1}{2p_-} \int_0^{l}\! d\sigma \, \pi_i \pi_i =
              \frac{p_-}{2l}\sum_{n=-\infty}^{\infty} p^i_n p^i_n
\end{equation}

For periodic boundary conditions
$Z^{\,i}(\sigma,\tau) = Z^{\,i}(\sigma+l,\tau)$
(closed string), a general solution to the equations
of motion can be written as
\begin{eqnarray}
\label{eq:AdS-Closed-Mode}
    &&Z^j(\sigma,\tau) = \sum_{n=-\infty}^{\infty} \left(
                       z^j_n + \frac{p^j_n}{l} \tau \right)
                       e^{\textstyle{\frac{2\pi i n \sigma}{l}}} \ , \\
    &&\pi_j = p_- \dot Z^j \,=\, p_- \!\sum_{n=-\infty}^{\infty}
                       \frac{p^j_n}{l} \;
                       e^{\textstyle{\frac{2\pi i n \sigma}{l}}}
\end{eqnarray}
Reality of $Z^i(\sigma,\tau)$ requires
$z^i_{-n} = (z^i_n)^*$, $p^i_{-n} = (p^i_n)^*$.
The periodicity of $Z^-$ does not follow
from the constraint: relation (\ref{eq:Zminusprime})
tells one only that $Z^{-\prime}$ is periodic;
$Z^-$ itself might have a term linear in $\sigma$,
which violates periodicity.
Thus, we must impose periodicity
of $Z^-$ as an extra requirement.%
\footnote{
     This can be viewed as fixing the last gauge freedom of
     $\sigma$ shifts by a constant (which is present only in
     the closed string case).
     In quantum theory, periodicity of $Z^-$ is equivalent to
     the condition that one works in the subspace of the
     Hilbert space which has total momentum along the closed
     string equal to zero.}
The mode expansion for $Z^-(\sigma,\tau)$ again has an analogous form:
\begin{equation}
    Z^- (\sigma,\tau) = \sum_{n=-\infty}^{\infty} \left(
                        z_n^- + \frac{p_n^-}{l} \tau \right)
                        e^{\textstyle{\frac{2\pi i n \sigma}{l}}}
\end{equation}
and the coefficients are given by the same formulas
(\ref{eq:small-z-minus}), (\ref{eq:small-p-minus}).
As in the open string case, $z_0^-$ is undetermined.
Note that $z_{-n}^- = (z_n^-)^*$, $p_{-n}^- = (p_n^-)^*$.
The periodicity of $Z^-$ becomes the condition
$n z_n^- \big|_{n=0} = 0$, which translates to
\begin{equation}
    \sum_{m=-\infty}^{\infty} m \, z^i_{-m} \, p^i_m = 0 \ .
\end{equation}
The mode expansion of the Hamiltonian is:
\begin{equation}
\label{eq:closed string H mode expansion}
    H_{LC} \; = \;
             \frac{p_-}{2l}\!\sum_{n=-\infty}^{\infty} p^i_n p^i_{-n}
             \; = \;
             \frac{p_-}{2l}\!\sum_{n=-\infty}^{\infty} p^i_n (p^i_n)^* \ .
\end{equation}

In the gauge where distance along
the string is measured in units of $|\pi_Y|$,
the description of dynamics looks different.
The independent degrees of freedom now are
${\bm Z}^\perp \equiv Z^{i=2..d-1}$,
${\bm \pi}_\perp$, $Z^-$, and $\pi_-$,
while $Y$ is determined by the constraints
\begin{eqnarray}
    && \dot Y =
       \frac{p_Y}{\pi_-} \ ,\\
    && Y' = -\frac{
       \pi_- Z^{-\prime} +
       {\bm \pi}_\perp \cdot
       {\bm Z}^{\perp\prime} }{p_Y}  \ .
\end{eqnarray}
The constraints fix $Y$ only up to an
overall additive constant, and therefore
the zeromode $y_0$ of $Y(\sigma,\tau)$ is
left undetermined, analogously to $z^-_0$
in the discussion of the $\pi_- = const$ gauge.

\subsection{String quantization}

To quantize, one imposes canonical
commutation relations on the
independent variables
(including the unconstrained zeromodes).
In the $\pi_- = const$ gauge,
\begin{eqnarray}
    &&\left[ Z^j (\tau, \sigma), \pi_k (\tau, \sigma') \right] =
    i \delta(\sigma - \sigma') \, \delta^j_k
        \label{eq:comm-rels-transverse-fields}
\end{eqnarray}
For 
the closed string, commutation relations
(\ref{eq:comm-rels-transverse-fields})
give
%
%
%
%
\begin{equation}
    [z^j_n, p^k_m] = \frac{i}{p_-} \delta^{jk} \delta_{n,-m}
\end{equation}
For 
the open string 
we will take
$p^i_{-n}=p^i_n$, $z^i_{-n}=z^i_n$,
and treat only mode expansion coefficients
with $n\geq 0$ as independent degrees of freedom.
With the normalization of the $p^i_n$s above,
the mode expansion coefficients satisfy
\begin{equation}
\label{eq:open-zp-comm}
    \left[ z^j_n , \, p^k_m \right] =
    \frac{i}{2p_-} \delta_{nm} \delta^{jk}
    , \ \ \ \ n,m\geq 0
\end{equation}
For both open and closed strings,
\begin{equation}
[z_0^-, p_-] = i/l \ ,
\end{equation}
and both
$p_-$ and $z_0^-$
commute with all
transverse mode coefficients.
The Hamiltonians 
(\ref{eq:open string H mode expansion}),
(\ref{eq:closed string H mode expansion})
contain
only momenta, and therefore have
no ordering ambiguity.
The eigenstates of the Hamiltonians
have positive energy, and thus there
is no tachyon in the spectrum.

In the $\pi_Y = const$ gauge, one has
\begin{eqnarray}
    &&\left[ Z^j (\sigma,\tau), \pi_k (\sigma',\tau) \right] =
    i \delta(\sigma - \sigma') \, \delta^j_k \ ,
    \ \ \ \ \ j,k = 2..d-1 \ , \\
    && \left[ Z^- (\sigma), \pi_- (\sigma')\right] =
    i \delta(\sigma - \sigma') \ .
\end{eqnarray}
The zeromode of the fixed momentum
$\pi_Y$ commutes with all independent
degrees of freedom, but does not
commute with the zeromode of $Y$:
\begin{equation}
\label{eq:py commutation relation}
     \left[ y_0, p_Y \right] = i/l \ .
\end{equation}

\subsection{Cutoff in the warped direction}

So far we saw that single-string
Hamiltonians of string theory on
AdS space exactly reproduce the
Hamiltonians of free scalar field
theory in the sector of single-trace
states with a large number of particles.
Let us now show that a non-zero mass of
particles in the field theory can be
viewed as coming from a cutoff in
the $Y$ direction, that is, from the
requirement that $Y$ is finite-ranged.
This is very similar to the supergravity
construction of confining gauge theories
in the strong coupling limit of the AdS/CFT
correspondence.

In that limit, one finds the following picture.
A fundamental string, whose ends on the boundary
represent external source particles in field
theory, is gravitationally pulled away from
the boundary, in the direction of increasing $Y$.
Since the string can not move past the cutoff,
the potential energy between the external
sources is linearly proportional to their
separation at large distances.
Alternatively, one can solve the wave equation
for the corresponding supergravity field
with the boundary conditions that the field
vanishes at the cutoff and is normalizable
at the boundary.
In this case the boundary conditions make the
spectrum of the wave equation discrete, and
the eigenvalues are interpreted as masses of
the particles created by the operator which
couples to the corresponding supergravity field.
An analogous situation takes place in the
zero-coupling limit we consider here.

To see this, it is convenient to work
in the $\pi_Y=const$ gauge.
Let us impose the cutoff in the $Y$
direction, so that $0<Y<Y_{max}$.
The $Y$-momentum zeromode commutes with both
${\bm \pi}_\perp$ and $\pi_-$, and therefore
in the Schr\"odinger equation with Hamiltonian
(\ref{eq:py string Hamiltonian}),
the $y_0$-dependent part of the wave function
can be factored out.
The commutation relation
(\ref{eq:py commutation relation})
implies that $p_Y$ has a simple
representation as
$p_Y = -i/l\;\partial/\partial y_0$,
and therefore one has to solve
\begin{equation}
\label{eq:Schroedinger equation}
    -\frac{1}{l^2} \frac{d^2}{dy_0^2} \psi(y_0) =
     E_Y \psi(y_0) \ ,
\end{equation}
where $0<y_0<Y_{max}$, and $E_Y$ is the
eigenvalue of $p_Y^2$.
By analogy with the strong-coupling case
we impose Dirichlet boundary conditions
$\psi(y_0 = 0) = \psi(y_0 = Y_{max}) = 0$,
which gives
\begin{equation}
    E_Y = \frac{4\pi^2 n^2}{l^2 Y_{max}^2} \ ,
\end{equation}
where $n=0,1,2\dots$ is the mode number.
Thus for the eigenstates
(\ref{eq:Schroedinger equation})
the string Hamiltonian
(\ref{eq:py string Hamiltonian})
becomes
\begin{equation}
  H_{LC} = \int_0^l d\sigma \, \frac{1}{2\pi_-(\sigma)}
           \left(
                     \bm{\pi}_{\perp}(\sigma)^2 +
                     \left(\frac{2\pi}{Y_{max}}\right)^2
                     \frac{n^2}{l^2}
           \right)
\end{equation}
This is to be compared to the continuum
version of the field theory Hamiltonian
with constant density of particles,
(\ref{eq:FT continuum Hamiltonian 1}),
in the limit when the number of particles
$M$ is large:
\begin{equation}
    H_{LC} = \int_0^l dx\; \frac{1}{2 \, \pi_- (x)}
    \left(\pip^2 (x) + m^2 \frac{M^2}{l^2} \right) \ .
\end{equation}
Thus, one can identify the mass of the
free scalar as
\begin{equation}
    m = \frac{2\pi}{Y_{max}} \ ,
\end{equation}
and the number of particles $M$ of the
field theory state as the mode number $n$
of the zeromode wavefunction.
Massless field theory gives rise to
string theory on the AdS space with
cutoff removed, $Y_{max}\to\infty$.
The freedom to impose the cutoff on the AdS space is the
freedom to add a mass term to the free
scalar.

\section{Critical dimension of the bosonic string}
\label{sec:critical dimension}

In this section we argue that the
string theory which one obtains by
taking the naive zero-radius limit
of the classical Hamiltonian
(\ref{eq:LC AdS Hamiltonian})
does not have a critical dimension,
as expected from its relation to the
free scalar field theory.

\subsection{Critical dimension in light-cone gauge}

The critical dimension of string
theory arises from the fact that Weyl
invariance of the Polyakov action
becomes anomalous unless coupled
to a matter CFT with the correct
central charge.
In light-cone gauge this anomaly
is subtle to detect: Weyl and
reparametrization invariance get
completely fixed, and it is
spacetime Lorentz invariance that
picks up the anomaly.
Rotations in the transverse space
are  manifest in light-cone gauge.
Not manifest are the rotations which
take $Z^1$ (the spatial coordinate
singled out to form $Z^{\pm}$), to
one of the transverse $Z^i$s.
Such rotations have to be accompanied
by a compensating
reparametrization/rescaling
of the world-sheet.
It is the new $Z^+$ that has to be equal
to $\tau$ after the rotation.
A failure of reparametrization/rescaling
to be a good symmetry shows up as an
anomaly in the Lorentz algebra.

This story is well known in flat space~\cite{GSW}.
A similar argument can be made for the
case  of AdS space.
With the AdS metric written in Poincare
coordinates (\ref{eq:AdS metric}), the
Lorentz invariance of the action
(\ref{eq:string action})
(rotations of $Z^{\mu=0..d-1}$)
is manifest.
For this Lorentz subgroup of the AdS
isometry group, the standard
N\"other procedure gives rise to the
conserved charges
\begin{equation}
    J^K_I = \int_0^l d\sigma \, \left(
            Z^K \pi_I -
            \eta_{I\!J} \, \eta^{KL} Z^J \pi_L
            \right) \
\end{equation}
(large Latin indices run from $0$ to $d-1$).
By imposing the canonical
commutation relations
(\ref{eq:comm-rels-transverse-fields}),
one finds the usual Lorentz algebra
\begin{equation}
\label{eq:Lorentz-algebra}
    [J^K_I, J^M_N] = i \left(
                     \delta^K_N J^M_I -
                     \delta^M_I J^K_N +
                     \eta_{N\!I}\,\eta^{M\!L} J^K_L -
                     \eta_{N\!L}\,\eta^{M\!K} J^L_I
                     \right) \ .
\end{equation}
The algebra implies that
the light-cone components
\begin{equation}
    J^M_\pm = \frac{1}{\sqrt{2}}\left(J^M_1 \pm J^M_0 \right) =
             \int_0^l d\sigma \left(
              Z^M \pi_\pm - Z^\mp \pi_L \delta^{LM} \right)
\end{equation}
must satisfy
\begin{equation}
\label{eq:JpJp-zero}
    [J^M_+, J^N_+] = 0 \ .
\end{equation}
A non-zero value for the commutator (\ref{eq:JpJp-zero})
signals the presence of the conformal anomaly.
In principle it is possible that
by some accident the Lorentz algebra
will be anomaly free while
conformal symmetry will exhibit some
anomaly.
This is, however, unlikely,
since the anomaly is really a failure
of diffeomorphisms times Weyl invariance, and the
Lorentz generators we are analyzing do
require the compensating transformations
on the world-sheet.

\subsection{Scaling the anomaly away}

As we saw in Section
$\ref{sec:choice of sigma}$,
taking the $\alpha' \to \infty$ limit
of the classical string Hamiltonian
in flat space gives the AdS Hamiltonian
in the limit $R^2/\alpha'\to 0$.
Of course, taking $\alpha'$ to infinity
is not a very sensible thing to do, because
$\alpha'$ is a dimensionful quantity.
All this limit means is that we are
studying states on time scales very
short compared to $\alpha'$.
On the other hand, $R^2/\alpha'$ is
a dimensionless parameter,
and taking it to zero in the classical
string Hamiltonian on AdS makes perfect
sense.
Still, it follows that there is a
limit, in which the usual flat space
free string spectrum reduces to the
zero radius AdS spectrum by
taking $\alpha' \rightarrow \infty$,
while suitably rescaling $\tau$.
We show below that in this limit the
usual flat-space anomaly scales to zero.
All mode expansions will be written
in the $\pi_- = const$ gauge.

For open strings,
a general solution to the flat space
equations of motion
is given by
\begin{equation}
    Z^m(\sigma, \tau) = \left(z^m_0 + \frac{p^m_0}{l} \tau \right)
                        \; + \;
                        i\sqrt{2} \sum_{k \neq  0} \frac{a^m_k}{k}
                        \exp\left[-\frac{i \pi k c \tau}{ l}\right]
                        \cos\left( \frac{\pi k \sigma}{l} \right) .
\end{equation}
where $c = \frac{1}{2 \pi \alpha' p_-}$
is the dimensionless velocity of the
wave propagating along the string.
Canonical commutation relations
(\ref{eq:comm-rels-transverse-fields})
imply that the flat space mode expansion
coefficients satisfy
\begin{equation}
\label{eq:flat-mode-comm}
   [a_k^m, a_{-k}^n] =
   \alpha' k\delta^{mn} \sim\frac{1}{c}\delta^{mn} \,.
\end{equation}
Now we can take the
$\alpha' \rightarrow \infty$
(that is, $c\rightarrow 0$) limit.
Expanding out the exponential,
keeping the constant and the linear
terms, and comparing with the AdS result
(\ref{eq:AdS-Open-Mode})
\begin{equation}
     Z^m(\sigma, \tau) =
           \left(z^m_0 + \frac{p^m_0}{l } \tau \right) \; + 2 \;
           \sum_{k > 0} \left(z^{m}_k + \frac{p^{m}_k}{ l} \tau \right)
           \cos\left( \frac{\pi k \sigma}{l} \right) \ ,
\end{equation}
we find
\begin{eqnarray}
     z^m_k &=& \frac{i}{\sqrt{2}}
               \left(\frac{a^m_k}{k}-\frac{a^m_{-k}}{k}\right)\ ,
     \label{eq:AdS z in terms flat modes open}\\
     p^m_k &=& \frac{\pi c}{\sqrt{2}}\left(a^m_k + a^m_{-k} \right) \,.
     \label{eq:AdS p in terms flat modes open}
\end{eqnarray}
Keeping these two quantities
finite while sending
$\alpha'\to\infty$ reduces
the flat space solution to the
zero radius AdS solution.
That is, we take $a^m_k$
(rather, their eigenvalues) to
infinity, while keeping
their differences finite.

The flat space solution
for closed string is
\begin{equation}
Z^m(\sigma, \tau) = (z^m_0 + \frac{p^m_0}{l} \tau) \; + \;
 \frac{i}{\sqrt{2}} \sum_{k \neq  0}\left\{
   \frac{a^m_k}{k}\exp\left[\frac{2\pi i k (\sigma - c\tau)}{l}\right]
             \; + \;
   \frac{\tilde{a}^m_k}{k}\exp\left[-\frac{2\pi i k (\sigma + c\tau)}{l}\right]
  \right\} \,,
\end{equation}
Again, comparing this with the
AdS result for the closed string
(\ref{eq:AdS-Closed-Mode})
\begin{equation}
     Z^m(\sigma, \tau) = \left(z^m_0 + \frac{p^m_0}{l} \tau \right)
            \; + \;
            \sum_{k \neq  0}\left(z^m_k + \frac{p^m_k}{l}\tau \right)
            e^{\textstyle{\frac{2\pi i k \sigma}{l}}} \ ,
\end{equation}
we obtain
\begin{eqnarray}
     z_k^m &=& \frac{i}{\sqrt{2}}\left(\frac{a^m_k}{k}-
               \frac{\tilde{a}^m_{-k}}{k}\right)\,,\\
     p_k^m &=& \sqrt{2}\pi c(a^m_k + \tilde{a}^m_{-k}) \,.
\end{eqnarray}

Now it is easy to see that the anomaly vanishes.
The commutator of interest is
\begin{equation}
     [J^m_+,J^n_+]_{AdS}=
     \lim_{c\rightarrow 0}[J^m_+,J^n_+]_{flat}\,.
\end{equation}
The Lorentz generators $J^m_+$
in flat space also have the same
form as the ones in AdS,
$$
    J^m_+ = \int_0^l d\sigma \left(
              Z^m \pi_+ - Z^- \pi_l \delta^{lm} \right) \,.
$$
The constraint equation
(\ref{eq:LC-constraint1})
implies that
$\pi_+$ is proportional to $c^2$,
while $Z^-$ contains a factor of $c$. Hence the
commutators have the generic form
$$
  [J_+,J_+]_{flat} \sim c^4 [aaa,aaa]\,.
$$
Consider the open string case for the concreteness.
In flat space, the anomaly arises from
the terms quadratic in $a^m_k$s (see e.g. \cite{GSW}). When one reduces
the commutator $[aaa,aaa]$ to $aa$, the factor of $1/c^2$ is introduced
by the commutation relation (\ref{eq:flat-mode-comm}). Thus the
anomaly scales as
\begin{eqnarray}
  [J^m_+,J^n_+]_{AdS}&=&\lim_{c\rightarrow 0}[J^m_+,J^n_+]_{flat}\sim
        \lim_{c\rightarrow 0}\sum_{k=1}^{\infty}c^2\Delta_k
              (a^m_{-k}a^n_k - a^n_{-k}a^m_k) =\\
        &=&\lim_{c\rightarrow 0}
            \sum_{k=1}^{\infty}c
            \frac{ik}{\pi}\Delta_k(z^m_kp^n_k-z^n_kp^m_k)=0 \,,
\end{eqnarray}
where we used
(\ref{eq:AdS z in terms flat modes open}), 
(\ref{eq:AdS p in terms flat modes open})
in the penultimate step, and
$$\Delta_k = k\left(\frac{26-(d+1)}{12}\right) +
\frac 1k \left(\frac{(d+1)-26}{12} +2 (1-a) \right)$$
is an $\alpha'$-independent number \cite{GSW}.
Here $a$ is undetermined
normal ordering constant in $Z^-$,
which does not enter the Hamiltonian.
We see that in the limit $c \to 0$
Lorentz algebra closes for any values
of $d$ and $a$.

The closed string case works similarly.
Thus we conclude that there
is no critical dimension.

\section{Discussion}
\label{sec:discussion}

We have shown that states with many particles in a large $N$ free
scalar field theory are well described in terms of a string
world-sheet. The string theory is tensionless in the sense that it
has no $\sigma$ derivatives. Thus the string can fall apart into
``partons''. Still, it is described by a Hamiltonian which can be
obtained from a non-linear sigma model on a quite singular target
space by light-cone gauge fixing, and it exhibits $\sigma$
reparametrization invariance.

An interesting question is of course how to turn on interactions.
There are two types of interactions of interest. For one we can
look at finite $N$ corrections. Already in the free scalar theory
the combinatorics of the Wick contractions of the free fields give
a non-trivial $1/N$ expansion. In particular only in the large
$N$ limit does one get a decoupling of single- and multi-trace
states. The $1/N$ corrections correspond to the $g_s$ expansion of
string theory, allowing strings to split and join. On the field
theory side the theory is solvable to all orders in $N$. One can
take this again as a definition of the interacting string theory.
One might then wonder if (at least for states with a large number
of partons), this expansion can be reobtained from purely stringy
reasoning, e.g. from light-cone string field theory.

More interesting would be to turn on interactions in the field
theory. Since the Hamiltonians of the two theories are the same,
any such perturbation can be mapped to a deformation of the bulk
theory. One interesting question is whether there is any coupling
in the field theory that would correspond to turning on finite
curvature radius in the bulk. Since the string perturbation involves
$(Z')^2$ terms, whose matrix elements can become arbitrarily large, 
for a generic string
theory this perturbation will not make sense. This is
consistent with the fact that non-supersymmetric conformal field
theories are hard to find. One might hope that in the case of type
IIB string theory on $AdS_5$ $\times$ $S^5$ and its dual ${\cal
N}=4$ supersymmetric Yang-Mills theory this perturbation is well defined. 
After all, we know that in the ${\cal N}=4$ SYM theory
the coupling can be turned on smoothly.
In fact,
while we were finishing this work,~\cite{Dhar Mandal Wadia} appeared, 
which, using
basically the same ideas in the supersymmetric context, presents
evidence that this perturbation is indeed valid.

Last but not least, another very interesting application of our
results is the problem of closed string tachyon condensation. Flat
space is an unstable solution of bosonic or type 0 string theory.
It has been a long standing puzzle to determine what happens when
the tachyon rolls down the hill. One natural problem is
that the potential energy decreases, so the most naive guess would
be that the true minimum should have a negative cosmological
constant. Our zero radius AdS solutions would be the natural
candidate for the endpoint of tachyon condensation in the
bosonic string. So our speculation is that as the closed string
tachyon condenses, the potential for the tachyon becomes arbitrarily
negative, while the dilaton gets frozen at $1/g_s \sim e^{-\Phi}\sim N$ for some
integer $N$. Very similar backgrounds should
exist in type 0 string theory, with the dual theory being 
a theory of free vector
particles.

\acknowledgments
\noindent We would like to thank Allan Adams, Tom Banks, Simeon
Hellermann, Ami Katz, Steve Pinsky, Eva Silverstein, and Charles Thorn for
helpful discussions. This work was partially supported by the DOE
under contract DE-FG03-96-ER40956.


\end{document}